\documentclass[12pt,notitlepage,onecolumn]{article}
\usepackage{amsmath}
\usepackage{harvard}

\input{tcilatex}

\begin{document}

\title{Counterpropagating Wavepacket Solutions of the Time-Dependent Schr%
\"{o}dinger Equation for a Decaying Potential Field}
\author{Babur M. Mirza\thanks{%
E-mail: bmmirza2002@yahoo.com} \\
Department of Mathematics \\
Quaid-i-Azam University, Islamabad. 45320 Pakistan.}
\maketitle

\begin{abstract}
We investigate wavepacket solutions for time-dependent Sch\"{o}dinger
equation in the presence of an exponentially decaying potential. Assuming
for travelling wave solutions the phase to be a linear combination of the
space and time coordinates, we obtain two distinct wavepacket solutions for
the Schr\"{o}dinger equation. The wavepackets counterpropagate in space at a
constant velocity without any distortion or spreading thus retain their
initial form at arbitrarily large distances.
\end{abstract}

\section{Introduction}

Wavepacket solutions of the Schr\"{o}dinger equation and their
interpretation has been investigated for diverse cases of physical interest.
In non-relativistic quantum theory attempts have been made to find
wavepacket solutions for such potential fields as the Morse and the
generalized Morse potentials [1,2], harmonic and pseudoharmonic potentials
[3], the Woods-Saxon potentials [4,5], and in the presence of a Coulomb
field [6]. In these cases a complete correspondence with the classical
theory can be established in general [7], although for time-dependent
potentials this correspondence is usually more complicated and in most cases
not possible in the sense of Ehrenfest theorem [8]. A persistent feature of
wavepackets, both in case of time independent as well as time dependent
potentials, is their spatial spreading with time. Propagating free
wavepackets based on Fourier construction methods exhibit this spreading
property especially. However there are cases, such as the free Bessel
wavepacket solutions [9-11], where the spreading does not occur. These
wavepackets have been demonstrated in lab (see for example [12,13] and
references therein). In general nonspreading Bessel wavepackets are not
normalizable so that the Ehrenfest theorem cannot applied.

In this paper we investigate new exact wavepacket solutions of the Sch\"{o}%
d-inger equation in the presence of an exponentially decaying potential
field. Such potentials are of interest due to the fact that exponentially
decaying fields are easily set up and maintained in practice. In the
asymptotic limit an exponentially decaying potential becomes constant hence
the wavepacket can be considered as free at a sufficiently large distance
from the source. Also the exponentially decaying potential considered here
can be reduced to some special cases of time-independent potentials, such as
the generalized Morse potential or the inverse Wood-Saxon potential. \ We
solve the time-dependent Schr\"{o}dinger equation by splitting it into the
quantum Hamilton-Jacobi equation, and the continuity equation. We assume a
propagating wave solution by considering a linear combination of the space
and time variables for the phase function. We thus obtain two possible
wavepacket solutions, one for the time variable $t$ and another with the
time inversion $-t$ propagating in the opposite directions. These solutions
are expressed as the zeroth order Bessel function, and posses the
nonspreading property.

The paper is organized as follows. In section 2 we solve the time-dependent
Schr\"{o}dinger equation for the potential\ using the corresponding quantum
Hamilton-Jacobi and the continuity equations. Travelling wave form of the
solutions is then obtained by requiring that the phase function is a linear
combination of the space and time coordinates. In section 3 we discuss the
physical properties of the wavepacket solutions. It is shown that the
wavepackets can be considered as a continuous superposition of travelling
waves that propagate in opposite directions at each instance of time with a
constant velocity. It is also observed that the wavepackets retain their
shape throughout their motion without any dissipation. In the last section
we summarize the main results of the study.

\section{The One Dimensional Bessel Wavepackets}

We consider the decaying potential of the form $Ce^{-x-kt}$ where $k$ is a
dimensional constant later fixed. The potential function rapidly decreases
as $x$ increases so that at a sufficiently large distance the particle can
be considered as free. The time-dependent Schr\"{o}dinger equation is then
given as%
\begin{equation}
-\frac{\hbar ^{2}}{2m}\frac{\partial ^{2}\Psi (x,t)}{\partial x^{2}}=i\hbar 
\frac{\partial \Psi (x,t)}{\partial t}+Ce^{-x-kt}\Psi (x,t).
\end{equation}%
Let the solution $\Psi (x,t)$ be given as%
\begin{equation}
\Psi (x,t)=R(x,t)e^{iS(x,t)/\hbar }.
\end{equation}%
This leads to the equation for the real part: 
\begin{equation}
-\frac{\partial S}{\partial t}=\frac{1}{2m}\left( \frac{\partial S}{\partial
x}\right) ^{2}+Ce^{-x-kt}-\frac{\hbar ^{2}}{2m}\left( \frac{1}{R}\frac{%
\partial ^{2}R}{\partial x^{2}}\right) ,
\end{equation}%
where the last terms is the quantum potential. Corresponding to the
imaginary part we have the continuity equation%
\begin{equation}
\frac{\partial R}{\partial t}+\frac{1}{m}\left( \frac{\partial R}{\partial x}%
\right) \left( \frac{\partial S}{\partial x}\right) +\frac{1}{2m}\left( R%
\frac{\partial ^{2}S}{\partial x^{2}}\right) .
\end{equation}%
For a travelling wave solution we assume that the phase function $S(x,t)$ is
expressed as $\alpha (x-\beta t)$ where $\alpha $ and $\beta $ are
constants. This gives in equation (3):%
\begin{equation}
\frac{\alpha ^{2}}{2m}+Ce^{-x-kt}-\frac{\hbar ^{2}}{2m}\left( \frac{1}{R}%
\frac{\partial ^{2}R}{\partial x^{2}}\right) =\alpha \beta .
\end{equation}%
Denoting the constant $a=(\alpha ^{2}/2m)-\alpha \beta $ we have the
differential equation%
\begin{equation}
\frac{\partial ^{2}R}{\partial x^{2}}=\frac{2m}{\hbar ^{2}}\left(
Ce^{-x-kt}+a\right) R.
\end{equation}%
This equation can be solved by separating the variables for $R(x,t)$, which
gives%
\begin{equation}
R(x,t)=C_{1}J_{-q}\left( \sqrt{\frac{8mC}{\hbar ^{2}}}e^{-(x+kt)/2}\right)
\Gamma (1-q)+C_{2}J_{q}\left( \sqrt{\frac{8mC}{\hbar ^{2}}}%
e^{-(x+kt)/2}\right) \Gamma (1+q),
\end{equation}%
where $q=\sqrt{-2am/\hbar }$, $J_{q}(z)$ is the $qth$ order Bessel function
and $\Gamma (z)$ is the gamma function.

We now consider equation (4). With the phase function $\alpha (x-\beta t)$
we have%
\begin{equation}
\frac{\partial R}{\partial t}=-\frac{\alpha }{m}\left( \frac{\partial R}{%
\partial x}\right)
\end{equation}%
Equation (8) is satisfied by $R(x,t)$ given in equation (7) provided that $%
q=0$, $\alpha =2m$ and $k=-2$. Alternatively equation (8) holds for $q=0$,
when $\alpha =-2m$ and $k=2$. We thus have for the $\Psi (x,t)$ function two
solutions:%
\begin{equation}
\Psi _{1}(x,t)=cJ_{0}\left( \sqrt{\frac{8mC}{\hbar ^{2}}}e^{-(x-2t)/2}%
\right) e^{2mi(x-t)/\hbar }\text{, when }k=-2;
\end{equation}%
and%
\begin{equation}
\Psi _{2}(x,t)=cJ_{0}\left( \sqrt{\frac{8mC}{\hbar ^{2}}}e^{-(x+2t)/2}%
\right) e^{-2mi(x-t)/\hbar }\text{, when }k=2;
\end{equation}%
where $c=C_{1}+C_{2}$ is a constant. Both $\Psi _{1}$ and $\Psi _{2}$
satisfy the Sch\"{o}dinger equation in the limiting case of a free particle
that is $V=0$. Although $\Psi _{1}$ and $\Psi _{2}$ are two distinct
solutions of the time-dependent Sch\"{o}dinger equation, a linear
superposition of these does not satisfy (1). Notice that with $k=\pm 2$ the
potential is $Ce^{-x\pm 2t}$.

\section{Physical Properties of the Wavepacket Solutions}

The integral representation of the Bessel function%
\begin{equation}
J_{0}\left( \sqrt{\frac{8mC}{\hbar ^{2}}}\exp (\pm t-\frac{x}{2})\right) =%
\frac{1}{\pi }\int_{0}^{\pi }\cos (\sqrt{\frac{8mC}{\hbar ^{2}}}\exp (\pm t-%
\frac{x}{2})\sin t)dt
\end{equation}%
shows that the wavepackets can be interpreted as a continuous superposition
of plane wave solutions $\cos (\sqrt{8mC/\hbar ^{2}}\exp (\pm t-\frac{x}{2}%
)\sin t)$.

Figure (1) and (2) show the probability densities corresponding to the $\Psi 
$ function solutions (9) and (10) at \ different time instances. We notice
that the wavepackets propagate in opposite directions. The propagation speed
can be calculated using $v=(\partial S/\partial x)/m.$ This shows that both
wavepackets counterpropagate with equal velocity. Another feature of the
wavepacket solutions is nonspreading (exhibited in figure (3) and (4) in
three dimensions). This property is retained by the wavepacket even at very
large distances where the effects of the potential function are negligible.
Thus in the asymptotic limit both wavepackets are free, and \ move in
opposite directions while retaining their initial shapes.

\section{Conclusions}

In this paper we have obtained wavepacket solutions of the Sch\"{o}dinger
equation in the presence of an exponentially decaying potential field. Two
coupled differential equations, namely the Hamilton-Jacobi equation and the
equation of continuity, were obtained by making a substitution of the form $R
$ $e^{iS/\hbar }$. These equation lead to two independent solutions
representing continuous superpositions of travelling plane waves
collectively forming a wavepacket. These wavepackets counterpropagate with a
constant velocity and posses the property of nonspreading hence retain their
form at very large distances.

\bigskip

FIGURE\ CAPTIONS:

Figure 1: Probability density\ plots for the wavepacket $\Psi _{1}$ at
different times with $8mC/\hbar ^{2}=1$, exhibiting nonspreading while
propagating in the negative $x$-direction.

Figure 2: Probability density\ plots for the wavepacket $\Psi _{2}$ at
different times with $8mC/\hbar ^{2}=1$, exhibiting nonspreading while
propagating in the positive $x$-direction.

Figure 3: 3D plot for the probability density for one dimensional Bessel
wavepacket $\Psi _{1}$ with $8mC/\hbar ^{2}=1$.

Figure 4: 3D plot for the probability density for one dimensional Bessel
wavepacket $\Psi _{2}$ with $8mC/\hbar ^{2}=1$.

\bigskip

\end{document}